\documentclass[aps,prb,preprint,groupedaddress,showpacs]{revtex4-1}
\usepackage{latexsym}
\usepackage{amsfonts}
\usepackage[english]{babel}
\usepackage[dvips]{epsfig}
\usepackage{natbib}
 \usepackage{xcolor}[2007/01/21]
\usepackage[dvipdfm,bookmarks=true,bookmarksnumbered,bookmarksopen]{hyperref}
\hypersetup{pdfpagemode=FullScreen,colorlinks=true,
citecolor=green,unicode,bookmarksopenlevel=3,
pdftoolbar=true,pdfauthor={Helen Gomonay},pdftitle={Solid oxygen}}
\begin{document}

\title{Magnetic structures of $\delta$-O$_2$ resulting from competition of interplane exchange interactions}
\author{E. V. Gomonay$^{1}$, V. M. Loktev$^{2}$}
\affiliation{~$^{1}$National Technical University of Ukraine
``KPI'', ave Peremogy, 37, 03056 Kyiv, Ukraine
\\ ~$^{2}$ Bogolyubov Institute for Theoretical Physics NAS of Ukraine,\\ Metrologichna str. 14-b, 03680, Kyiv,
Ukraine\\ E-mail: {\rm Helen Gomonay} $\langle{\tt
malyshen@ukrpack.net}\rangle$, {\rm Vadim Loktev} $\langle{\tt
vloktev@bitp.kiev.ua}\rangle$ }

\begin{abstract}
Solid oxygen is a unique molecular crystal whose phase diagram is
mostly imposed by magnetic ordering, i.e., each crystal phase has
a specific magnetic structure. However, recent experiments showed
that high-pressure $\delta$-phase is implemented in different
magnetic structures. In the present paper we study the role of
interplane exchange interactions in formation of the magnetic
structures with different stacking sequences of the close-packed
planes. We show that temperature-induced variation of
intermolecular distances 
 can give rise to compensation of
the exchange coupling between the nearest close-packed planes and
result in the phase transition between different magnetic
structures within $\delta$-O$_2$. Variation of the magnetic
ordering is, in turn, accompanied by the step-wise variation of
interplane distance governed by space and angular dependence of
interplane exchange constants.
\end{abstract}
\maketitle
\section{Introduction}


Solid oxygen is known to occupy a particular place in the large
family of cryocrystals. First, it is the only molecular crystal
that shows magnetic ordering in a wide range of temperatures and
pressures \cite{Freiman:2004}. On the other hand, some magnetic
modifications of solid O$_2$ have recently found a practical
application as converters for the production of ultra-cold
neutrons \cite{Frei:2010un, Lavelle:2010}.

\begin{figure}[htbp]
 \centering{\includegraphics[width=0.7\columnwidth]{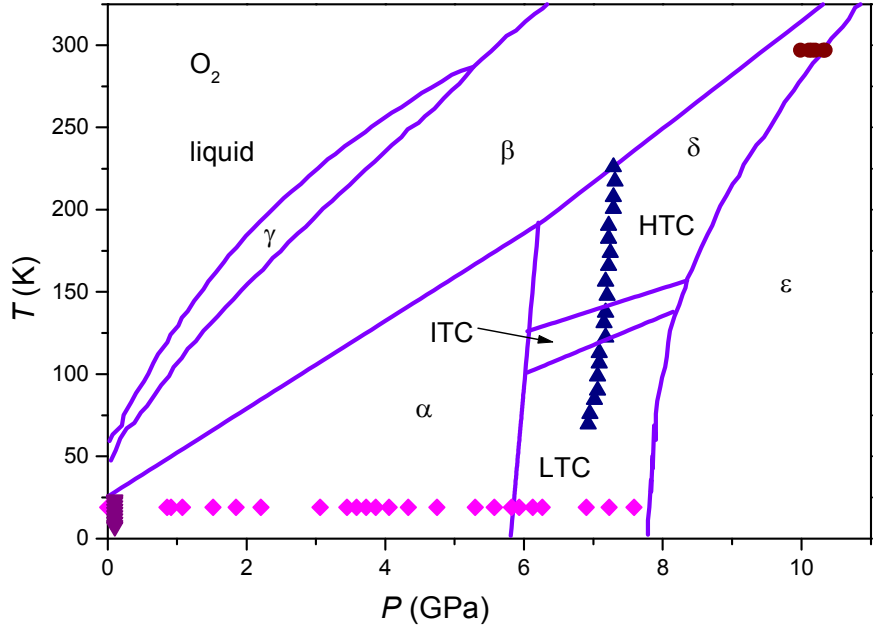}
\caption{(Color online) Phase diagram of oxygen. Solid lines show
the generally accepted phase boundaries \cite{Klotz:2010}. Points
indicate the experimental data used for calculations of
intermolecular distances shown in Fig.~\ref{fig_l_vs_v}:  isobaric
trajectories are represented by up triangles \cite{Klotz:2010}
($P\approx 7.3$~GPa, $\delta$-phase) and down triangles
\cite{Krupskii:1979E} (ambient pressure, $\alpha$-phase),
isotherms are represented by diamonds \cite{Akahama:2001}
($T=19$~K, $\alpha$-phase) and circles \cite{Olinger:1984} (RT,
$\delta$-phase). HTC, ITC and LTC denote, correspondingly, high,
intermediate and low temperature commensurate magnetic structures
\cite{Klotz:2010}.}\label{fig_phase_diagram}}
\end{figure}

Due to magnetic properties of O$_2$ molecule that possesses
nonzero spin $S_{O_2}=1$ in the ground state, solid oxygen shows
rich and nontrivial phase diagram that includes, among others,
different magnetic phases ($\alpha$, $\beta$, $\delta$, and
$\varepsilon$, see Fig.\ref{fig_phase_diagram}). Exchange magnetic
interactions between O$_2$ molecules  at low temperature prove to
be of the same order as lattice energy. As a result, the phase
diagram of the solid oxygen is completely imposed by the magnetic
structures, i.e., the crystal structure is being ``spin
controlled'' \cite{Goncharenko:2004}. However, recent experiments
by Klotz et al \cite{Klotz:2010} revealed the different types of
magnetic ordering within the high-pressure $\delta$-phase,
nontrivial temperature behavior of the lattice constants, and put
into doubt the dominant role of magnetic interactions.

In the present paper we try to corroborate the idea of
spin-controlled crystal structure of solid oxygen. We argue that
temperature dependence of lattice parameters in $\delta$-phase
results from variation of the inter- and intra-plane exchange
magnetic coupling. We show that competition between the different
interplane exchange constants induced by the lattice deformation can
generate a variety of the magnetic structures with different
stacking sequence of the close-packed $ab$ planes.



\section{Model}
\begin{figure}[htbp]
 \centering{\includegraphics[width=0.5\columnwidth]{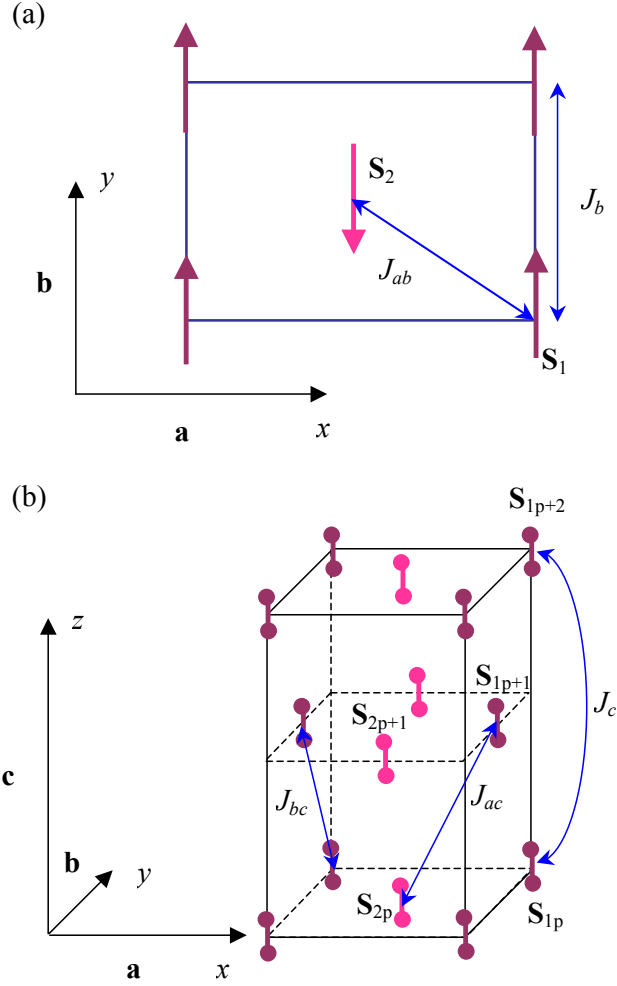}
\caption{(Color online) The choice of magnetic sublattices and
exchange coupling constants in $\delta$-O$_2$: (a) $\mathbf{S}_1$,
$\mathbf{S}_2$ are spin vectors of two sublattices within
$ab$-plane directed nearly along the $\mathbf{b}$-axis, $J_{b}$
and $J_{ab}$ are intra- and intersublattice exchange constants,
correspondingly; (b) the choice of sublattices in different
$ab$-planes is arbitrary, spins $\mathbf{S}_{1p}$ and
$\mathbf{S}_{1p+1}$ could be either parallel or antiparallel
depending on the magnetic structure (see Fig.~\ref{fig_3D}).
Parameters $J_{ac}$ and $J_{bc}$ describe interplane exchange
coupling between the nearest neighboring $ab$-planes, and $J_{c}$
describes next-to-nearest neighbor interplane exchange coupling
along $\mathbf{c}$ direction.}\label{fig_xyplane}}
\end{figure}

Crystal structure of $\delta$-O$_2$ is described by orthorhombic
symmetry group \cite{Freiman:2004} D$_{2h}^{23}$ . The oxygen
molecules can all be considered to be oriented parallel to each
other and perpendicular to the close-packed atomic $ab$-planes
\cite{Akahama:2001}, as shown in Fig.~\ref{fig_xyplane}(b). As it
as already mentioned, each O$_2$ molecule has a spin $S_{O_2}=1$
in its ground state that determines magentic properties of solid
oxygen. The magnetic ordering within $ab$-plane corresponds to
collinear antiferromagnet (AFM) and is described by two magnetic
sublattices \cite{Goncharenko:2004}, $\mathbf{S}_1$ and
$\mathbf{S}_2$
 (see Fig.~\ref{fig_xyplane}a). It is worth
noting that the in-plane ordering is similar in $\alpha$- and
$\delta$-phases with the magnetic moments directed nearly along
the $\mathbf{b}$-axis.
In what follows we suppose the in-plane AFM configuration
unchangeable, in accordance with the experimental data
\cite{Goncharenko:2004}. The mutual orientation of the moments in
the adjacent close-packed layers is not uniquely determined and
can be parallel or antiparallel, as will be shown below.

Temperature dependence of the lattice constants and magnetic phase
diagram of $\delta$-O$_2$ could be explained from the analysis of
the magnetic energy of the crystal (per unit volume) which in the
mean-field approximation takes a form
\begin{eqnarray}\label{Hamiltonian}
w_{\rm mag}
&=&\frac{1}{N}\sum_p\left[2J_{b}(r_b)\left(\mathbf{S}_{1p}^2+\mathbf{S}_{2p}^2\right)+4J_{ab}(r_{ab})\mathbf{S}_{1p}\mathbf{S}_{2p}\right.\nonumber\\
&+&\left.J_{bc}(r_{bc})\left(\mathbf{S}_{1p}\mathbf{S}_{1p+1}+\mathbf{S}_{2p}\mathbf{S}_{2p+1}\right)+J_{ac}(r_{ac})\left(\mathbf{S}_{1p}\mathbf{S}_{2p+1}+\mathbf{S}_{2p}\mathbf{S}_{1p+1}\right)\right.\nonumber\\
&+&\left.J_{c}(r_c)\left(\mathbf{S}_{1p}\mathbf{S}_{1p+2}+\mathbf{S}_{2p}\mathbf{S}_{2p+2}\right)\right].
\end{eqnarray}
Here the vectors
 $\mathbf{S}_{\alpha p}$ ($\alpha=1,2$) are the spins averaged over the $p$-th $ab$-plane, $N$ is the number of $ab$-planes per unit
 length, different constants $J(r)$  describe the
 in-plane and interplane exchange interactions between the nearest and
next to the nearest neighbors (NN and NNN) separated by a distance
$r$ (as shown in Fig.~\ref{fig_xyplane}b), $r_b=b$,
$r_{ab}=\sqrt{a^2+b^2}/2$, $r_{bc}=\sqrt{b^2+c^2}/2$,
$r_{ac}=\sqrt{a^2+c^2}/2$, and $r_c=c$, vectors $\mathbf{a},
\mathbf{b}$, and $\mathbf{c}$ define the orthorhombic unit cell.
All the spins have the same value $|\mathbf{S}_{\alpha
 p}|=M_0(T)$ which is supposed to be temperature dependent.


 We assume that the exchange coupling between O$_2$ molecules
has an AFM character (all the exchange constants are positive,
$J>0$). Basing on the analysis made in
Ref.\onlinecite{Freiman:2004} we further assume that the in-plane
exchange integrals $J(r)$ are the decreasing functions of
intermolecular distances $r$, so, $dJ(r)/dr<0$. The interplane
exchange integrals $J(r, \theta)$ are, in addition, the functions
of angle $\theta$ between the molecular axes and intermolecular
vector $\mathbf{r}$ (see, e.g., Refs.\onlinecite{Wormer:1983,
Wormer:1984, Uyeda:1985}).

 According to experimental data
\cite{Klotz:2010}, variation of lattice parameters $a$, $b$ and
$c$ within the wide temperature range is small and thus can be
described by the components $u_{jj}\ll 1$ ($j=x,y,z$) of the
strain tensor as follows:
\begin{equation}\label{strain}
  a=a_0(1+u_{xx}),\quad b=b_0(1+u_{yy}),\quad c=c_0(1+u_{zz}),
\end{equation}
where $a_0, b_0, c_0$ are the lattice parameters at $T\rightarrow
0$ (for a fixed pressure value), and coordinate axes $x,y,z$ are
parallel to the axes of the orthorhombic crystal unit cell (see
Fig.~\ref{fig_xyplane}). In what follows we introduce three
combinations of $u_{jj}$ that form irreducible representations of
the space group D$_{2h}^{23}$:
 \textit{i}) relative variation of the specific volume $\delta v/v\equiv
  u_{xx}+u_{yy}+u_{zz}$;
   \textit{ii}) rhombic deformation of in-plane unit cell $u_{\rm rh}\equiv
  u_{xx}-u_{yy}$;
 \textit{iii}) variation of interplane distance $u_{zz}$.

Elastic energy written in these notations takes a form
\begin{equation}\label{elastic}
  w_{\rm el}(\hat
  {u})=\frac{1}{2}c_{\rm rh}
  u_{\rm rh}^2+\frac{1}{2}c_{33}u_{zz}^2+f\left(\frac{\delta
  v}{v};
  T\right)+P\frac{\delta v}{v},
\end{equation}
where $c_{\rm rh}$, $c_{33}$ are elastic modula, $T$  and $P$ are
temperature and external pressure correspondingly, $f(\delta v/v;
T)$ is a model function that takes into account
temperature-induced anharmonicity of the crystal lattice.

Magnetoelastic contribution into free energy of the crystal is
obtained from (\ref{Hamiltonian}) by expansion of the exchange
parameters $J(r)$ in series over small strains $u_{jj}$ as it was
done, e.g., in Ref.\onlinecite{Loktev:1981E} (see below).

\section{Qualitative considerations}
\begin{figure}[htbp]
 \centering{\includegraphics[width=0.5\columnwidth]{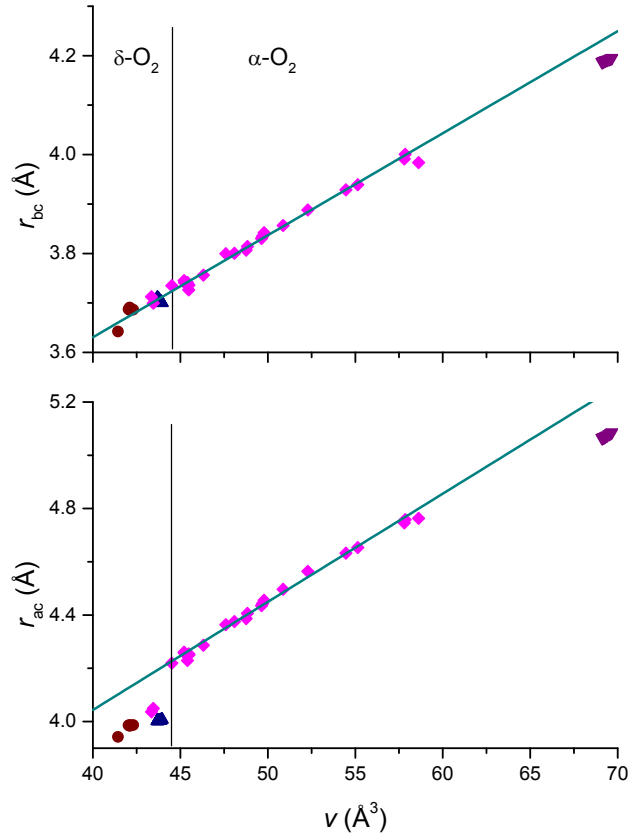}
\caption{(Color online) Volume dependence of intermolecular
distances $r_{bc}$, $r_{ac}$. Experimental data are taken from
Refs.\onlinecite{Klotz:2010} (up triangles),
\onlinecite{Krupskii:1979E} (down triangles),
\onlinecite{Akahama:2001} (diamonds), \onlinecite{Olinger:1984}
(circles), see also Fig.\ref{fig_phase_diagram}. Lines show linear
approximation calculated for $\alpha$-phase.}\label{fig_l_vs_v}}
\end{figure}

We argue that the observed temperature variation of the crystal
and magnetic properties of $\delta$-O$_2$ arises from competition
of the AFM exchange interactions between different sites and
proceed from the following.
\renewcommand{\theenumi}{\arabic{enumi}}
\renewcommand{\labelenumi}{\theenumi.}
\begin{enumerate}
  \item Experiments \cite{Klotz:2010} demonstrate negative thermal
  expansion along $\mathbf{b}$ and positive thermal expansion along $\mathbf{a}$-axes.
  This fact can be explained by competition between the in-plane exchange constants $J_{b}$ and $J_{ab}$ (see
  Fig.~\ref{fig_xyplane}a). The quantity $J_{b}$ couples the spins with
  parallel orientation and thus gives rise to an increase of the magnetic energy, while $J_{ab}$ couples the antiparallel
  spins and gives rise to the energy decrease. Energy growth due to temperature
  variation of
  $M_0(T)$ can be compensated by the negative thermal expansion of
  $r_b$ (effective repulsion) and positive thermal expansion of $r_{ab}$
  (effective attraction) that means that thermal expansion
  along $\mathbf{a}$ direction should be larger than contraction along $\mathbf{b}$ direction.
 \item Magnetic phase diagram of $\delta$-O$_2$ includes three
 phases with different stacking sequences of $ab$-planes (see
 Fig.~\ref{fig_3D}). This fact can be explained by competition of
the  interplane exchange interactions $J_{bc}$, $J_{ac}$, and
$J_{c}$ (see Fig.~\ref{fig_xyplane}b). It is evident that mutual
orientation of NN spins depends upon the sign of the difference
$J_{bc}-J_{ac}$. If $J_{bc}<J_{ac}$, the configuration with
$\mathbf{S}_{1p}\uparrow\uparrow \mathbf{S}_{2p+1}$  (labeled in
Ref.\onlinecite{Klotz:2010} as HTC phase\footnote{~HTC, ITC and
LTC mean high-, intermediate- and low temperature commensurate,
correspondingly.}) is energetically favorable. In the opposite
case $J_{bc}>J_{ac}$ the configuration
$\mathbf{S}_{1p}\uparrow\uparrow \mathbf{S}_{1p+1}$ (LTC phase) is
more favorable. If, for some reasons, $J_{bc}\approx J_{ac}$, an
equilibrium configuration is governed only by relatively small NNN
exchange interactions, i.e., by $J_{c}>0$, and corresponds to
antiparallel coupling of $\mathbf{S}_{1p}$ and $\mathbf{S}_{1p+2}$
spins (ITC phase). It should be noticed that the idea that ``the
interactions between the third interplane neighbours can stabilize
the ferromagnetic coupling of O$_2$ planes even if all the
exchange constants are ... antiferromagnetic'' was advanced by
Goncharenko et al \cite{Goncharenko:2004}, before the magnetic
structure of $delta$-phase was ultimately established.

\item Both $\alpha$- and $\delta$-phases have the same magnetic
ordering within $ab$-plane. However, $\delta$-O$_2$ shows a variety
of the magnetic structures with different stacking sequences of
$ab$-planes, while $\alpha$-O$_2$ shows only one stacking sequence
(corresponding to HTC $\delta$-O$_2$) in the whole range of
temperature and pressure values. This fact can also be explained by
competition of the interplane exchange interactions $J_{bc}$ and
$J_{ac}$. To clarify this point we have plotted the intermolecular
distances $r_{ac}$
  and
$r_{bc}$ as the functions of average intermolecular distances
represented by the volume $v$ of the crystal unit cell (see
Fig.~\ref{fig_l_vs_v}). The distances $r_{ac}$
  and
$r_{bc}$ were calculated using the results of measurement of
temperature \cite{Klotz:2010, Krupskii:1979E} and pressure
\cite{Akahama:2001, Olinger:1984} dependencies taken in different
regions of solid O$_2$ phase diagram including $\alpha$- and
$\delta$-phases\footnote{~Correspondence between NN in $\alpha$-
and $\delta$-phases was established from the assumption that
$\alpha$-phase transforms into $\delta$-phase by the continuous
shift of the close-packed planes \cite{gomo:2005}}. It can be
clearly seen that in the $\alpha$-O$_2$ the distance
$r_{ac}>r_{bc}$ and the difference between these values is of the
order of 20\%. Taking into account the character of space and
angular dependence $J(r, \theta)$, one can assume that
$J_{bc}<J_{ac}$ and HTC ordering is energetically favorable. At
the $\alpha\rightarrow\delta$ transition point the dependence
$r_{ac}(v)$ shows step-like decrease. Correspondingly, the
relative difference between $r_{ac}$ and $r_{bc}$ diminishes down
to $\propto 7.5\%$. Corresponding difference  between $J_{bc}$ and
$J_{ac}$ can be compensated due to strong angular dependence of
$J(r, \theta)$ that becomes crucial at small intermolecular
distances. Thus, the difference $J_{bc}-J_{ac}$ changes sign and
ITC and LTC phases turn out to be favorable.
\end{enumerate}
In the next sections we will substantiate these qualitative
considerations with the phenomenological analysis of the magnetic
and  crystal structure of $\delta$-O$_2$.

\section{Intraplane exchange and temperature dependence of lattice parameters}
Equilibrium values  of lattice parameters $a$, $b$ and $c$ at
given temperature and pressure  are calculated from minimization
of free energy  $w=w_{\rm mag}+w_{\rm el}$ (see
Eqs.~(\ref{Hamiltonian}) and (\ref{elastic})) with respect to
parameters $\delta v/v$, $u_{\rm rh}$ and $u_{\rm zz}$. In the
first approximation we neglect small contribution of interplane
exchange \cite{Klotz:2010}, $J_{ac}(\propto J_{bc})/J_{ab}<1/30$.
Intraplane exchange constants depend on the deformations
implicitly, through the intermolecular distances $r(\delta v/v,
u_{\rm rh}, u_{\rm zz})$, see. e.g. Ref.\onlinecite{Loktev:1981E}.
We further assume that the thermal-expansion
coefficient\footnote{~Induced by lattice anharmonicity only,
without magnetic contribution.}, $\beta_v$, and isothermal
compliance $\chi_T$ are constant in the considered part of phase
diagram, so, the function $f$ in equation (\ref{elastic}) can be
written as \cite{gomo:2005}:
\begin{equation}\label{function}
  f\left(\frac{\delta
  v}{v};
  T\right)=\frac{1}{2\chi_T}\left(\frac{\delta
  v}{v}\right)^2-\frac{\beta_vT}{\chi_T}\left(\frac{\delta
  v}{v}\right).
\end{equation}

In this case equilibrium values of deformations at a given AFM
magnetic structure are the following:
\begin{equation}\label{v}
    \frac{\delta v}{v}=-\chi_T
    P+\beta_vT-2\chi_vM_0^2(T)\left(\frac{dJ_b}{dr}\left|_{r^{(0)}_{b}}\right.-\frac{dJ_{ab}}{dr}\left|_{r^{(0)}_{ab}}\right.\right),
\end{equation}
\begin{equation}\label{u_rh}
    u_{\rm rh}=\frac{2M_0^2(T)}{c_{\rm
    rh}}\left(\frac{dJ_{b}}{dr}\left|_{r^{(0)}_{b}}\right.+\frac{a_0^2-b_0^2}{a_0^2+b_0^2}\frac{dJ_{ab}}{dr}\left|_{r^{(0)}_{ab}}\right.\right),
\end{equation}
and
\begin{equation}\label{u_zz}
    u^\perp_{zz}=\frac{2M_0^2(T)}{c_{33}}\left(\frac{dJ_b}{dr}\left|_{r^{(0)}_{b}}\right.-\frac{dJ_{ab}}{dr}\left|_{r^{(0)}_{ab}}\right.\right).
\end{equation}
Superscipt ``$\perp$'' in Eq.~(\ref{u_zz}) indicates intraplane
exchange contribution into $u_{zz}$.

\begin{figure}[htbp]
 \centering{\includegraphics[width=0.7\columnwidth]{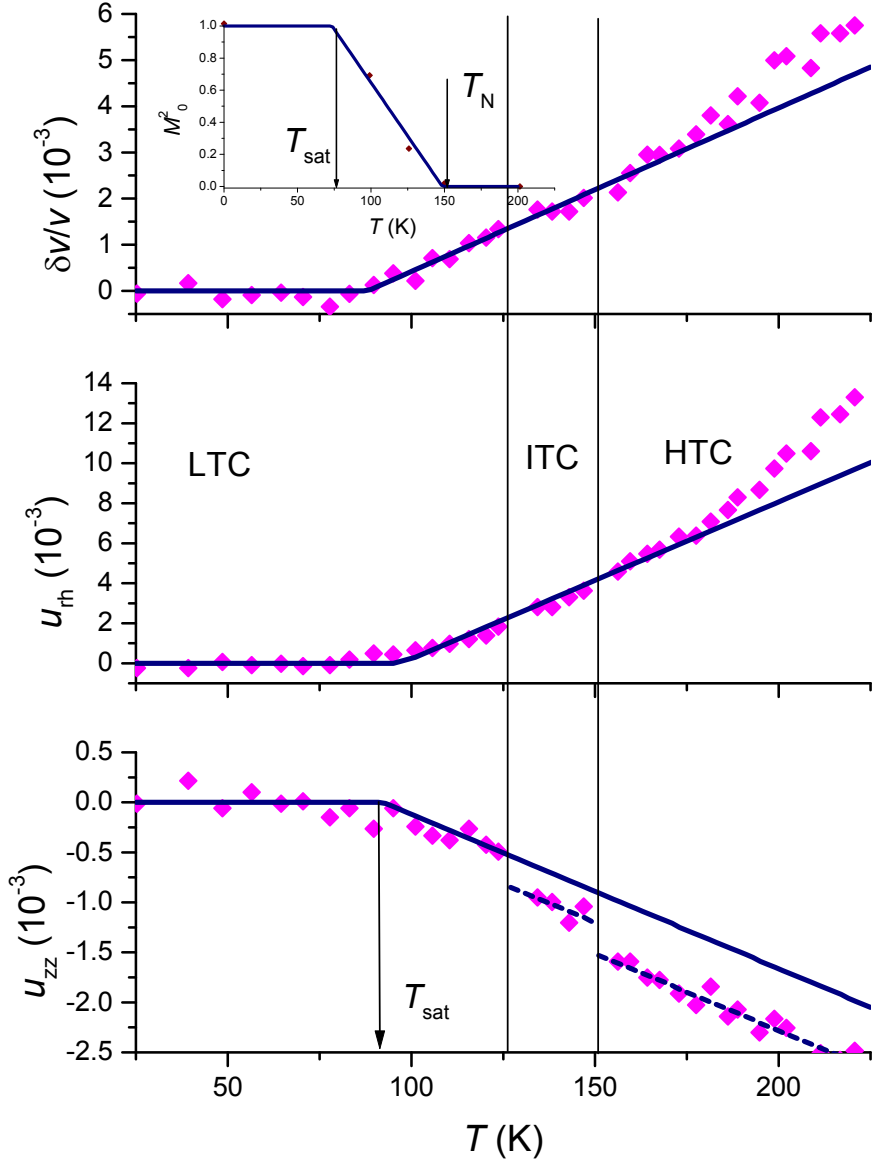}
 \caption{(Color online) Temperature dependence of the parameters $\delta v/v$
 (upper panel), $u_{\rm rh}$ (middle panel) and $u_{zz}$ (bottom
 panel). Points -- experimental data \cite{Klotz:2010}, solid
 lines -- theoretical approximation according to
 Eq.~(\ref{g-function}). Inset shows temperature dependence of
 $M_0^2(T)$ (normalized to 1) taken at $P=7.6$~GPa: points -- experimental data
 \cite{Goncharenko:2005}, solid line -- approximation according to
 Eq.~(\ref{M-function}).}\label{fig_defor_vs_T}}
\end{figure}

Temperature dependence of the values $\delta v/v$, $u_{zz}$ and
$u_{\rm rh}$ can be unambiguously defined if we take into account
the following facts: \textit{i}) decreasing and exponential
character of $J(r)$ function; \textit{ii}) relations between
intermolecular distances at $T=0$: $r^{(0)}_{b}\equiv
b_0>r^{(0)}_{ab}\equiv\sqrt{a_0^2+b_0^2}/2$ and $a_0>b_0$;
\textit{iii}) temperature dependence of sublattice magnetization
$M_0(T)$ predicted by the mean-field theory and supported by
neutron diffraction measurements \cite{Goncharenko:2005} (see
inset in Fig.~\ref{fig_defor_vs_T}):
\begin{equation}\label{M-function}
  M^2_0(T)\propto\left\{\begin{array}{cc}
    const.,& T\le T_{\rm sat}, \\
    (1-T/T_N),  & T_{\rm sat}\le T\le T_N.
  \end{array}\right.
\end{equation}
Here $T_N$ is the N\'{e}el temperature and $T_{\rm sat}$ (usually
$\propto 0.5 T_N$) is the temperature at which $M_0$ attains its
saturation value.

As a result, cell volume, $\delta v/v$, and in-plane orthorhombic
deformation, $u_{\rm rh}$, are increasing functions of temperature
(because $dJ_b(r^{(0)}_{b})/dr<dJ_{ab}(r^{(0)}_{ab})/dr<0$), while
interplane distance (and corresponding deformation $u_{zz}$) is
decreasing function of temperature. All the dependencies could be
approximated with the function
\begin{equation}\label{g-function}
  g(T)=\left\{\begin{array}{cc}
    0,& T\le T_{\rm sat}, \\
    A(T-T_{\rm sat}),  & T_{\rm sat}\le T\le T_N,
  \end{array}\right.
\end{equation}
where the constant $A$ (1/K) depends upon the values $dJ/dr$.

 Fig.~\ref{fig_defor_vs_T}
shows the temperature dependencies of the deformations $\delta
v/v$, $u_{\rm rh}$, and $u_{zz}$ in $\delta$-O$_2$. Points
correspond to experimental data \cite{Klotz:2010}, solid lines are
approximations according to Eq.~(\ref{g-function}) with $T_{\rm
sat}=97$~K and
\begin{equation}\label{AB-constant}
  A=\left\{\begin{array}{ccc}
     3.54\cdot 10^{-5}&\textrm{for}&\delta v/v, \\
      7.85\cdot 10^{-5}& \textrm{for}&u_{\rm rh},\\
    -1.54\cdot 10^{-5}&\textrm{for}&u_{zz}.
  \end{array}\right.
\end{equation}
It can be clearly seen that temperature variation of the in-plane
deformation $u_{\rm rh}$ and  volume $\delta v/v$ can be
adequately explained by the temperature dependence of the in-plane
exchange interactions and, in particular, $M_0^2(T)$ (see
Eq.~(\ref{M-function})). As it was already mention, in-plane
exchange forces cause strong contraction along $r_{ab}$ direction
which gives dominant contribution into orthorhombic deformation
and volume effect. An analogous mechanism is responsible for the
anisotropic lattice compressibility within the $ab$-plane in a
wide range of pressure values, as was pointed out in
Ref.~\onlinecite{Nozawa:2008}.


However, interplane deformation $u_{zz}$ deflects from the
dependence (\ref{g-function}) in the transition points between the
magnetic phases with different stacking sequences of close-packed
planes (HTC, ITC and LTC). Full interpretation of the experimental
data is possible with the account of small but important
contribution of the interplane exchange interactions.





\section{Interplane exchange and different magnetic structures}
\begin{figure}[htbp]
 \centering{\includegraphics[width=0.7\columnwidth]{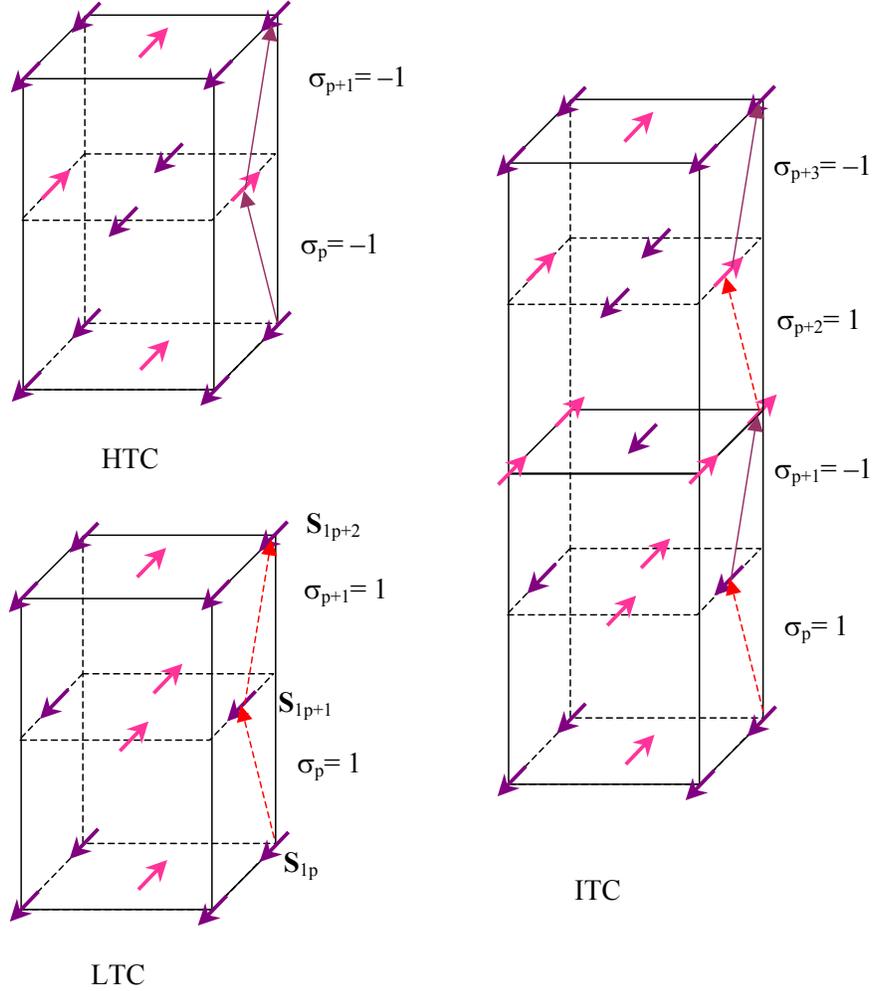}
\caption{(Color online) Three types of magnetic ordering in
$\delta$-phase. Left panel -- ``ferromagnetic ordering'' with the
order parameter
$\sigma_p\equiv\mathbf{S}_{1p}\cdot\mathbf{S}_{1p+1}/M_0^2 =-1$
(HTC, upper panel) or $\sigma_p=1$ (LTC, bottom panel). Right
panel -- ``antiferromagnetic'' ordering with the order parameter
$\sigma_p$ alternating between $\pm1$ from layer to layer
(ITC).}\label{fig_3D}}
\end{figure}

To elucidate the role of interplane exchange interactions in the
formation of equilibrium magnetic structures of $\delta$-O$_2$ we
reduce minimization of the magnetic energy (\ref{Hamiltonian}) to
the well-known (see, e.g., review Ref.~\onlinecite{Selke:1988})
1-dimensional ANNNI (Axial Next-Nearest Neighbor Ising) problem.
In assumption of the fixed in-plane spin ordering, the
3-dimensional magnetic structure is uniquely described by the set
of Ising variables (``pseudospins'') $\sigma_p=\pm 1$ defined as
follows (see also Fig.\ref{fig_3D}):
\begin{equation}\label{eta_definition}
  M_0^2\sigma_p\equiv\mathbf{S}_{1p}\mathbf{S}_{1p+1}=\mathbf{S}_{2p}\mathbf{S}_{2p+1}=-\mathbf{S}_{2p}\mathbf{S}_{1p+1}=-\mathbf{S}_{1p}\mathbf{S}_{2p+1}.
\end{equation}
Parameters $\sigma_p$ in fact define ``ferromagnetic'' (if
$\sigma_p=1$) or ``antiferromagnetic'' (if $\sigma_p=-1$) in two
neighboring ($p$-th and $p+1$-th) close-packed planes. Moreover,
if all the spins are collinear, mutual orientation of the
next-to-nearest neighboring planes is also defined by the same
parameters, e.g.,
$\mathbf{S}_{1p}\mathbf{S}_{1p+2}=(\mathbf{S}_{1p}\mathbf{S}_{1p+1})(\mathbf{S}_{1p+1}\mathbf{S}_{1p+2})/M_0^2=M_0^2\sigma_p\sigma_{p+1}$,
etc. (see Fig.~\ref{fig_3D}).

Thus, the magnetic energy (\ref{Hamiltonian}) in mean-field
approximation can be adequately presented in a form of
1-dimensional Ising model for the effective ``pseudospins''
$\sigma_p$:
\begin{equation}\label{Hamiltonian_ANNNI}
w_{\rm mag} =\frac{2M_0^2}{N}\sum_p \left[\Delta J_c\sigma_p
+J_{c}\sigma_p\sigma_{p+1}\right].
\end{equation}
It can be easily seen that the difference $\Delta J_c\equiv
J_{bc}-J_{ac}$ plays a role of the effective field that in the
absence of NNN coupling ($J_{c}=0$) tends to align all the
``pseudospins'' in parallel.  Such a ``ferromagnetic'' ordering
generates an LTC ($\sigma_p=1$, $\Delta J_c<0$) or HTC
($\sigma_p=-1$, $\Delta J>0$) magnetic structure (see
Fig.~\ref{fig_3D}). In turn, the exchange coupling between NNN,
$J_{c}$, is responsible for interaction between the neighboring
``pseudospins''. If $J_{c}<0$ (ferromagnetic exchange between the
``real'' spins), ``ferromagnetic'' coupling is still preferable
(LTC or HTC structures). However, if NNN exchange coupling is AFM,
$J_{c}>0$, then, an ``antiferromagnetic'' ordering of
``pseudospins'' ($\sigma_{2p}=1$,$\sigma_{2p+1}=-1$) that
corresponds to ITC structure is favorable.

Stability ranges of the HTC, ITC and LTC structures can be found
from comparison of corresponding energies:
\begin{equation}\label{energies}
w^{\rm LTC}_{\rm mag}=\left(-\Delta J_c+J_{c}\right)M_0^2, \quad
w^{\rm ITC}_{\rm mag}=-J_{c}M_0^2,\quad w^{\rm HTC}_{\rm
mag}=\left(\Delta J_c+J_{c}\right)M_0^2.
\end{equation}
Thus, HTC structure is stable if $\Delta J_c\le -2J_{c}$, ITC
structure is stable if $-2J_{c}\le\Delta J_c\le 2J_{c}$ and LTC
structure is stable if $2J_{c}\le\Delta J_c$. So, temperature
variation of interplane exchange coupling $J_{bc}$, $J_{ac}$, and
$J_{c}$ may generate a series of HTC-ITC-LTC phase transitions.

Now an important question is: ``What is the reason for variation
of the relation between the NN and NNN exchange constants in
$\delta$-O$_2$?''. We suppose that temperature variation of
interplane exchange constants is due to a strong and nontrivial
angular dependence of the exchange coupling. In particular,
\emph{ab initio}  calculations of the exchange interactions
between an isolated pair of O$_2$ molecules \cite{Wormer:1983,
Wormer:1984} revealed the following facts:  \textit{i}) exchange
coupling parameters $J(r, \theta)$ show nonmonotonic, strongly
oscillating behaviour as a function of angle $\theta$;
\textit{ii}) the values $\theta$ at which $J(r, \theta)$ attains
its minimal and maximal values are very sensitive to
intermolecular distance $r$; \textit{iii}) for a fixed
intermolecular distance $r$ the absolute value of $J(r, 0)$
(molecular axes are parallel to the intermolecular vector) is much
greater than $J(r, 90^\circ)$ (molecular axes are perpendicular to
the intermolecular vector); \textit{iv}) $J(r, \theta)$ is
oscillating around zero value for intermediate values of angles,
$\theta\propto20\div40^\circ$. On the other hand, experiment
\cite{Klotz:2010} gives $\theta_{\rm bc}=24.06^\circ$,
$\theta_{\rm bc}$ varies from $32.17^\circ$ in LTC to
$32.55^\circ$ in HTC structures and obviously $\theta_c=0$. Thus,
we conclude that \textit{i}) $\Delta J_c$ can change sign due to
the strong angular dependence of $J_{ac}$ that equates AFM and FM
exchange ($J_{ac}\approx J_{bc}$) at different ($r_{ac}>r_{bc}$)
distances and/or oscillation around zero value of both $J_{bc}$
and $J_{ac}$; \textit{ii}) the value of NNN coupling $J_{c}$ may
be comparable with $|\Delta J_c|$ because space relaxation of the
exchange constants for $\theta_c=0$ is compensated by the
enhancement due to angular dependence.

The hypothesis of strong space dependence of the interplane
exchange constants is also supported by the observed jumps of
interplane distance in the HTC-ITC and ITC-LTC transition points
(see Fig.\ref{fig_defor_vs_T}, lower panel): $u^{\rm
LTC}_{zz}-u^{\rm ITC}_{zz}=u^{\rm ITC}_{zz}-u^{\rm
HTC}_{zz}=2.9\cdot 10^{-4}$. Really, with account of interplane
exchange contribution the temperature dependence (\ref{u_zz}) of
$u_{zz}$ can be refined as follows:
\begin{equation}\label{uzz_refined}
  u_{zz}=u^\perp_{zz}+\frac{M_0^2(T)}{c_{33}}\left\{\begin{array}{ccc}
     (\lambda_1-\lambda_2)&\textrm{for}&\textrm{LTC}, \\
      \lambda_2& \textrm{for}&\textrm{ITC},\\
    (-\lambda_1-\lambda_2)&\textrm{for}&\textrm{HTC},
  \end{array}\right.
\end{equation}
where
\begin{eqnarray}\label{def_labda}
  \lambda_1&\equiv& \left[\frac{2c_0^2-a_0^2}{8r_{ac}^{(0)}}\left.\frac{\partial J_{ac}}{\partial r}\right|_0-\frac{2c_0^2-b_0^2}{8r_{bc}^{(0)}}\left.\frac{\partial J_{bc}}{\partial r}\right|_0\right]-\left[\frac{6r_{ac}^{(0)}}{c_0^2}\left.\frac{\partial J_{ac}}{\partial
  \theta}\right|_0-\frac{6r_{bc}^{(0)}}{c_0^2}\left.\frac{\partial J_{bc}}{\partial
  \theta}\right|_0\right],\nonumber\\
   &&\lambda_2\equiv\left|\frac{\partial J_c}{\partial
  r}\right|_0,
\end{eqnarray}
and subscript ``0'' denotes that arguments of $r$ and $\theta$ are
taken at $T\rightarrow 0$.

Thus, if $\lambda_1\gg \lambda_2$,  then $u^{\rm LTC}_{zz}-u^{\rm
ITC}_{zz}=u^{\rm ITC}_{zz}-u^{\rm HTC}_{zz}\approx
M_0^2\lambda_1/c_{33}>0$, in accordance with the experiment.
Comparison with experimental data makes it possible to estimate
space dependence of in-plane and interplane exchange constants
quantitatively:
\begin{equation}\label{comparison_1}
\left|\frac{u^{\rm LTC}_{zz}-u^{\rm
ITC}_{zz}}{u^\perp_{zz}}\right|=\left|\frac{dJ_{ac}/dr-dJ_{bc}/dr}{dJ_{b}/dr-dJ_{ab}/dr}\right|\propto
0.1.
\end{equation}
It is interesting to note that analogous increase of interplane
distances was also observed \cite{Akahama:2001} during the
pressure-induced transition from $\alpha$- to $\delta$-phase at
$T=19$~K. According to phase diagram (diamonds in
Fig.~\ref{fig_phase_diagram}), corresponding $\delta$-O$_2$ has a
LTC structure while $\alpha$-O$_2$ shows a HTC ordering, so,
interplane distance should be larger in $\delta$-O$_2$, as it is
predicted by (\ref{uzz_refined}).

\begin{figure}[htbp]
 \centering{\includegraphics[width=0.7\columnwidth]{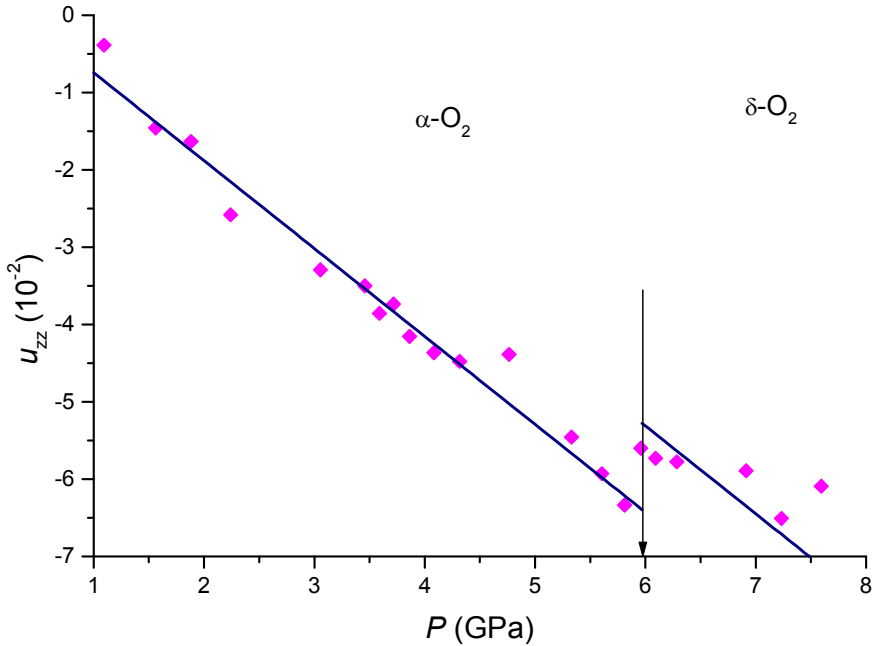}
 \caption{(Color online) Pressure dependence of $u_{zz}$. Points correspond to experimental data \cite{Akahama:2001} (see also Fig.~\ref{fig_phase_diagram}), solid lines are the best linear fit (see formula
 (\ref{linear_uzz_alpha})).}\label{fig_uzz_vs_P_alpha}}
\end{figure}

Fig.~\ref{fig_uzz_vs_P_alpha} shows the pressure dependence of
$u_{zz}$ calculated from experimental data
Ref.\onlinecite{Akahama:2001} (points) along with the linear
approximation according to formula
\begin{equation}\label{linear_uzz_alpha}
u_{zz}=-1.13\cdot 10^{-2} P +\left\{\begin{array}{ccc}
   0.38\cdot 10^{-2}, & \textrm{for}&\alpha-{\rm O}_2, \\
    1.45\cdot 10^{-2}, & \textrm{for}&\delta-{\rm O}_2.
     \end{array}\right.
\end{equation}
Assuming that pressure dependence $u_{zz}(P)$ results from the
space dependence of in-plane exchange constants (in analogy with
$u_{zz}(T)$) we get the same as (\ref{comparison_1}) estimation
for the in- and inter-plane exchange constants:
\begin{equation}\label{comparison_2}
\left|\frac{u^{\delta}_{zz}-u^{\alpha}_{zz}}{u_{zz}}\right|=\left|\frac{dJ_{ac}/dr-dJ_{bc}/dr}{dJ_{b}/dr-dJ_{ab}/dr}\right|\propto
0.15.
\end{equation}

\section{Conclusions}
In the present paper we have analyzed the role of interplane
exchange interactions in formation of the magnetic and crystal
structure of solid $\delta$-O$_2$. We show that the crystal volume
and orthorhombic deformation in $ab$-plane strongly depend on the
in-plane exchange forces. On the contrary, interplane distances
noticeably depend not only on the strong in-plane but also on
relatively small interplane exchange coupling. As a result, abrupt
change of the magnetic structure (HTC-ITC-LTC transition) is
followed by the step-wise variation of interplane distances.

We propose an interpretation of the mechanism of phase transitions
between the magnetic structures with different stacking sequence
of the $ab$-planes based on the competition between different,
relatively small interplane exchange integrals.
interpretation proposed in Ref.~\onlinecite{Klotz:2010} rests upon
assumption on strong temperature dependence of only one interplane
exchange constant $J_{bc}$ ($J_3$ in notations of
Ref.~\onlinecite{Klotz:2010}) induced by the libron excitations.
We argue that due to the strong angular and space dependence of
the exchange coupling the exchange forces between NN and NNN in
the $c$-direction could be of the same order value and should be
taken into account at the same foot. In this case the libron
contribution into all the exchange constants should be the same,
while configurational (i.e., depending on the relative positions
of molecules) contribution would be different. Correlation between
the experimental slope of the LTC-ITC-HTC and theoretical value
deduced in Ref.~\onlinecite{Klotz:2010} from the librational
fluctuations can be explained by the magnetic contribution into
librons parameters observed in Ref.~\onlinecite{Gorelli:2000}.

We supposed that the values of the exchange constants in solid
oxygen strongly depend upon the relative positions and orientation
of axes  of O$_2$-molecules. We proceeded from the calculations
\cite{Wormer:1983, Wormer:1984} for isolated pairs of molecules
that demonstrated an oscillatory character of $J(\theta)$
function. However, accurate values of the exchange constants for
certain configurations should be calculated with account of an
additional parameter, namely,  spatial orientation of
$\pi$-orbitals with respect to crystal axes. Such calculations are
beyond the scope of this paper.

In this paper we considered mainly the temperature effects that
cause variation of the crystal lattice parameters, interplane
exchange constants, and, as a result, series of transitions
between different magnetic phases. However, analogous effects
could be produced by pressure. Moreover, we assume that pressure
may induce some other than the considered commensurate magnetic
structures, especially in the vicinity of $\alpha-\delta$-
transition line.

\acknowledgments The authors would like to acknowledge Prof. Y. A.
Freiman for valuable assistance. The paper was partially supported
by the grants from Ministry of Science and Education of Ukraine
and Special Program of Fundamental Research of National Academy of
Sciences of Ukraine.



%
%
\end{document}